\documentclass[12pt]{iopart}
\usepackage{epsfig}
\eqnobysec
\newcommand{\Rref}[1]{reference~\cite{#1}}
\newcommand{\ba}{\begin{array}}
\newcommand{\ea}{\end{array}}
\newcommand{\eq}[2]{\begin{equation}#2\label{#1}\end{equation}}
\newcommand{\eqn}[1]{\begin{eqnarray}#1\end{eqnarray}}
\newcommand{\nn}{\nonumber}
\newcommand{\jour}[4]{\rm{#4}\ {\it #1}\ \bf{#2}\ \rm{#3}}
\newcommand{\book}[4]{\rm{#4}\ \it{#1}\ \rm{#2\ (#3)}}
\newcommand{\auth}[2]{\rm{#2}\ \rm{#1}}
\def\openone{\leavevmode\hbox{\small1\kern-3.3pt\normalsize1}}
\newcommand{\eps}{\epsilon}

\newcommand{\ep}{\varepsilon_p}
\newcommand{\ed}{\varepsilon_d}
\newcommand{\ves}{\varepsilon_s}
\newcommand{\ezx}{\varepsilon_{zx}}
\newcommand{\exy}{\varepsilon_{xy}}

\newcommand{\NCCO}{{\rm Nd}_{2-x}{\rm Ce}_x {\rm CuO}_{4-\delta}}

\newcommand{\cuo}{{\rm CuO}_2}
\newcommand{\ruo}{{\rm RuO}_2}
\newcommand{\cud}{{\rm Cu}3d_{x^2-y^2}}
\newcommand{\cus}{{\rm Cu}4s}
\newcommand{\ox}{{\rm O_a}2p_x}
\newcommand{\oy}{{\rm O_b}2p_y}
\newcommand{\tsp}{t_{sp}}
\newcommand{\tpd}{t_{pd}}
\newcommand{\tpp}{t_{pp}}
\newcommand{\tpdp}{t_{pd\pi}}
\newcommand{\tzz}{t_{zz}}
\newcommand{\tzzx}{t_{z,zx}}
\newcommand{\prl}{Phys. Rev. Lett.}
\newcommand{\PRB}{Phys. Rev. B}
\newcommand{\PC}{Physica C}
\newcommand{\PREP}{Phys. Rep.}
\begin{document}

\jl{3}
\title[Modelling of layered superconducting perovskites]{Tight-binding modelling of the electronic band structure of layered superconducting perovskites}
\author{T Mishonov\footnote[1]{Permanent address: Department of Theoretical Physics, Faculty of Physics,
University of Sofia, 5~J.~Bourchier~Blvd., 1164~Sofia, Bulgaria} and E Penev}
\address{Laboratorium voor Vaste-Stoffysica en Magnetisme, Katholieke Universiteit Leuven,\\ Celestijnenlaan 200 D, B-3001 Leuven, Belgium\\
\ \\
Received 9 March 1999, in final form 22 September 1999
}

\begin{abstract}
A detailed tight-binding analysis of the electron band structure of
the CuO$_2$ plane of
layered cuprates is performed within a $\sigma$-band Hamiltonian including four orbitals
 -- ${\rm Cu3}d_{x^2-y^2},{\rm Cu4}s, {\rm O2}p_x,$ and ${\rm O2}p_y.$
Both the experimental and theoretical hints in
favor of Fermi level located in a Cu or O band, respectively, are considered. For
these two alternatives analytical expressions are obtained for the LCAO electron
wave functions suitable for the treatment of electron superexchange. Simple
formulae for the Fermi surface and electron dispersions are derived by applying the L\"owdin down-fold procedure to set up the effective copper and oxygen Hamiltonians. They
are used to fit the experimental ARUPS Fermi surface of
Pb$_{0.42}$Bi$_{1.73}$Sr$_{1.94}$Ca$_{1.3}$Cu$_{1.92}$O$_{8+x}$ and both the
ARPES and LDA Fermi surface of $\NCCO$.
The value of presenting the hopping amplitudes as surface
integrals of {\it ab initio} atomic wave functions is demonstrated as well. The same approach is
applied to the RuO$_2$ plane of the ruthenate Sr$_2$RuO$_4.$ The LCAO Hamiltonians
including the three in-plane $\pi$-orbitals Ru4$d_{xy},$ O$_{\rm a}2p_y,$
O$_{\rm b}2p_x$ and the four transversal $\pi$-orbitals Ru4$d_{zx},$
Ru4$d_{yz},$ O$_{\rm a}2p_z,$ O$_{\rm b}2p_z,$ are separately considered. It is
shown that the equation for the constant energy curves and the Fermi contours
has the same canonical form as the one for the layered cuprates.
\end{abstract}

\pacs{74.25.Jb, 71.25.Pi, 74.72.-h}
\submitted{Published in J. Phys.: Condens. Matter \textbf{12} (2000), pp. 143--159}
\maketitle
%
%
\section{Introduction}
\label{sec:intro}

After the discovery of the high-$T_c$ superconductors the layered cuprates
became one of the most studied materials in the solid state physics. A vast range of
compounds were synthesised and their properties comprehensively investigated.
The electron band structure is of particular importance for understanding the
nature of superconductivity in this type of perovskites~\cite{Pickett}. Along this line 
one can single out the significant success achieved in the attempts to reconcile
the photoelectron spectroscopy data~\cite{Shen} and the band structure
calculations of the Fermi surface (FS) especially for compounds with
simple structure such as $\NCCO$~\cite{King,Yu91pr}.
A qualitative understanding, at least for the self-consistent electron picture, has been achieved
and for the most electron processes in the layered perovskites one can
employ adequate lattice models.

There is not much analysis of the electronic band structures of the high-$T_c$
materials in the terms of single analytical expressions available. This is
something for which there is a clear need, in particular to help in the
construction of more realistic many-body Hamiltonians.
The aim of this paper is
to analyse the common features in the electron band structure of the layered
perovskites within the tight-binding (TB) method~\cite{Bullett}. In the following we shall focus on the metallic (being eventually
superconducting) phase only, with the reservation that the antiferromagnetic
correlations, especially in the dielectric phase, could substantially change
the electron dispersions. It is shown that the linear combination of atomic
orbitals (LCAO) approximation can be considered as an adequate tool for analysing energy bands.
Within the latter exact analytic results are obtained for the constant energy
contours (CEC). These expressions are used to fit the FS of
$\NCCO$~\cite{King},
Pb$_{0.42}$Bi$_{1.73}$Sr$_{1.94}$Ca$_{1.3}$Cu$_{1.92}$O$_{8+x}$~\cite{Aebi},
and Sr$_2$RuO$_4$~\cite{Lu96} measured in angle-resolved
photoemission/angle-resolved ultraviolet spectroscopy (ARPES/ARUPS)
experiments.

In particular, by applying the L\"owdin perturbative technique for the $\cuo$
plane we give the LCAO wave function of the states near the Fermi energy
$\eps_{_{\rm F}}.$ These states could be useful in constructing the pairing
theory
for $\cuo$ plane. For the layered cuprates we find an alternative concerning
the Fermi level location---$\cud$ {\it vs.} O$2p\sigma$ character of the
conduction band. It is shown that analysis of extra spectroscopic data is needed
in order for this dilemma to be resolved.
As regards the $\ruo$ plane, the existence of three pockets of the FS
unambiguously reveals the Ru$4d\varepsilon$ character of the conduction
bands~\cite{Oguchi,Singh}.

To address the  conduction bands in the layered perovskites we start from a
common Hamiltonian including the basis of valence states O$2p$ and
Ru$4d\varepsilon,$ or $\cud,$ Cu$4s$ respectively for cuprates. Despite the
equivalent crystal structure of Sr$_2$RuO$_4$~\cite{Maeno} and
La$_{2-x}$Ba$_x$CuO$_4$~\cite{Bednorz}, the states in their conduction band(s)
are, in some
sense, complimentary. In other words, for the $\cuo$ plane the conduction band
is of $\sigma$-character while for the $\ruo$ plane the conduction bands are
determined by $\pi$ valence bonds. This is due to the separation into $\sigma$- and
$\pi$-part of the Hamiltonian $H=H^{(\sigma)}+H^{(\pi)}$ in first
approximation. The latter two Hamiltonians are studied separately.

Accordingly,
the paper is structured as follows. In \sref{sec:II} we consider the generic
$H^{(4\sigma)}$ Hamiltonian of the $\cuo$ plane~\cite{OKA94,OKA95} and
$H^{(\pi)}=H^{(xy)}+H^{(z)}$ is then studied in \sref{sec:III}. The results of
the comparison with the experimental data are summarised in \sref{sec:IV}.
Before embarking on a detailed analysis, however, we give an account of some clarifying issues concerning the applicability of the TB model and the band theory in general.

\subsection{Appology to the band theory}

It is well-known that the electron band theory is a self-consistent treatment of the electron motion in the crystal lattice. Even the classical 3-body problem demonstrates strongly correlated solutions, so it is {\it a priori} unknown whether the self-consistent approximation is applicable when describing the electronic structure of every new crystal. However, the one-particle band picture is an indispensable stage in the complex study of materials. It is the analysis of  experimental data using a conceptually clear band theory that reveals nontrivial effects: how strong the strongly correlated electronic effects are,  whether it is possible to take into account the influence of some interaction-induced order parameter back into the electronic structure etc. Therefore the comparison of the experiment with the band calculations is not an attempt, as sometimes thought, to hide the relevant issues---it is a tool to reveal interesting and nontrivial properties of the electronic structure.

Many electron band calculations have been performed for the layered perovskites and results were compared to data due to ARPES experiments. The shape of the Fermi surface is probably the simplest test to check whether we are on the right track or some conceptually new theory should be used from the very beginning. 

The tight-binding interpolation of the electronic structure is often used for fitting the experimental data. This is because the accuracy of that approximation  is often higher than the uncertainties in the experiment. Moreover, the tight-binding method gives simple formulae which could be of use for experimentalists to see how far they can get with such a simple minded approach.  The tight-binding parameters, however, have in a sense "their own life" independent of the {\it ab initio} calculations.
These parameters can be fitted directly to the experiment even when, by some reasons,  the electron band calculations could give wrong predictions. In this sense the tight-binding parameters are the appropriate intermediary between the theory and experiment. As for the theory, establishing of reliable one-particle tight-binding parameters is the preliminary step in constructing more realistic many-body Hamiltonians. The role of the band theory is, thus, quite ambivalent: on one hand, it is the final "language" used in efforts towards understanding a broad variety of phenomena; on the other hand, it is the starting point in developing realistic interaction Hamiltonians for sophisticated phenomena such as magnetism and superconductivity.  

The tight-binding method is the simplest one employed in the electron band calculations and it is described in every textbook in solid state physics; the layered perovskites are now probably the best investigated materials and the Fermi surface is a fundamental notion in the physics of metals. There is a consensus that the superconductivity of layered perovskites is related to electron processes in the CuO$_2$ and RuO$_2$ planes of these materials. It is not, however,  fair to criticise a given study, employing the tight-binding method as an interpolation scheme to the first principles calculations,  for not thoroughly discussing the many-body effects. The criticism should rather be readdressed to the {\it ab initio} band calculations. An interpolation scheme cannot contain more information than the underlying theory. It is not erroneous if such a scheme works with an accuracy high enough to adequately describe both the theory and experiment.   

In view of the above, we find it very strange that there are no simple interpolation formulae for the Fermi surfaces available in the literature and the experimental data are being published without an attempt towards simple interpretation. One of the aims of the present paper is to help interpret the experimental data  by the tight-binding method as well as setting up notions in the analysis of the {\it ab initio} calculations.
%
%
\section{Layered cuprates}
\label{sec:II}
%
%
\subsection{Model}
\label{sec:A}

The $\cuo$ plane appears as a common structural detail for all layered
cuprates. Therefore, in order to retain the generality of the considerations,  the electronic properties of the bare $\cuo$ plane will be addressed  without taking into account structural details  such as  dimpling, orthorhombic distortion, double
planes, surrounding chains etc. For the square unit cell with lattice constant $a_{_0}$ 
three-atomic basis is assumed $\{{\bf R}_{\rm Cu}, {\bf R}_{\rm O_b}, {\bf R}_{\rm O_b}\} = \{{\bf 0},(a_{_0}/2,0), (0,a_{_0}/2)\}.$
The unit cell is indexed by vector ${\bf n} = (n_x,n_y),$ where $n_x,$ $n_y=$ integer.
Within such an idealized model the LCAO wave function spanned over the
$|\cud\rangle,$ $|\cus\rangle,$ $|\ox\rangle,$ $|\oy\rangle$ states reads as
\eqn{
\fl \psi_{_{\rm LCAO}}( {\bf r} ) = \sum_{\bf n} \Bigl[ X_{\bf n}\psi_{_{ {\rm
  O_a}
  2p_x}} ({\bf r}-{\bf R}_{\rm O_a}-a_{_0}{\bf n}) +  Y_{\bf n} \psi_{_{ {\rm
O_b} 2p_y}} ({\bf r}-{\bf  R}_{\rm O_b}-a_{_0}{\bf n})\Bigr. \nn \\
 + \Bigl. S_{\bf n} \psi_{_{ {\rm Cu} 4s}}({\bf r}-{\bf R}_{\rm Cu}-a_{_0}{\bf
n}) + D_{\bf n}\psi_{_{{\rm Cu }3d}}({\bf r}-{\bf R}_{\rm Cu}-a_{_0}{\bf n})
\Bigr],
\label{LCAOwf}
}
where $\Psi_{\bf n}=(D_{\bf n},S_{\bf n},X_{\bf n},Y_{\bf n})$ is the
tight-binding wave function in lattice representation.

The neglect of the differential overlap leads to an LCAO Hamiltonian of the
$\cuo$ plane
\eqn{
 H & = & \sum\limits_{\bf n} \Bigl\{ D_{\bf n}^{\dagger} [ -t_{pd} (-X_{\bf
      n} + X_{x-1,y} + Y_{\bf n} - Y_{x,y-1}) + \eps_d D_{\bf n} ]
      \Bigr.\nn\\
  & & + S_{\bf n}^{\dagger} [ -t_{sp} (-X_{\bf n}+X_{x-1,y}-Y_{\bf n} +
         Y_{x,y-1}) + \eps_s S_{\bf n}] \nn\\
  & & + X_{\bf n}^{\dagger} [ -t_{pp} (Y_{\bf n}-Y_{x+1,y}-Y_{x,y-1}+
         Y_{x+1,y-1})                             \label{LCAOham} \\
  & &     \quad   - t_{sp}(-S_{\bf n}+S_{x+1,y})- t_{pd}(-D_{\bf
                   n}+D_{x+1,y}) + \eps_p X_{\bf n}] \nn \\
  & & + Y_{\bf n}^{\dagger} [ -t_{pp} (X_{\bf n}-X_{x-1,y}-X_{x,y+1}+
         X_{x-1,y+1}) \nn\\
  & &    \Bigl. \quad  - t_{sp}(-S_{\bf n}+S_{x,y+1})- t_{pd}(D_{\bf
      n}+D_{x,y+1}) + \eps_p Y_{\bf n}]\Bigr\},\nn
}
where the components of $\Psi_{\bf n}$ should be considered as being
Fermi operators. The notations $\epsilon_d,$ $\epsilon_s,$ and $\epsilon_p$
stand respectively for the  $\cud,$ $\cus$ and O$2p\sigma$
single-site energies. The direct $\ox \rightarrow \oy$ exchange is denoted by
$t_{pp}$ and similarly $\tsp$ and $\tpd$ denote the $\cus \rightarrow {\rm
O}2p$ and ${\rm O}2p \rightarrow \cud$ hoppings respectively. The sign rules for the
hopping amplitudes are sketched in~\fref{fig:1}---the bonding orbitals enter the
Hamiltonian with a negative sign. The latter follows directly from the surface
integral approximation for the transfer amplitudes, given in~\ref{ap1}.

For the Bloch states diagonalizing the Hamiltonian~(\ref{LCAOham})
\eq{Blochwf}{
\Psi_{\bf n} \equiv
\left(\!
 \ba{c}
  D_{\bf n} \\
  S_{\bf n} \\
  X_{\bf n} \\
  Y_{\bf n} \\
 \ea \!\right)
= \frac{1}{\sqrt{N}} \sum_{\bf p}
\left(
 \ba{c}
  D_{\bf p} \\
  S_{\bf p} \\
  \e^{{\rm i}\varphi_a}X_{\bf p} \\
  \e^{{\rm i}\varphi_b}Y_{\bf p} \\
 \ea \right)   \e^{{\rm i} {\bf p \cdot n}},
}
where $N$ is the number of the unit cells, we use the same phases as
in references~\cite{OKA94,OKA95}: $\varphi_a = \case12(p_x-\pi),$ $\varphi_b =
\case12(p_y-\pi).$
This equation describes the Fourier transformation between the coordinate representation
$\Psi_{\bf n}=(D_{\bf n}, S_{\bf n}, X_{\bf n}, Y_{\bf n}),$ with $\bf n$ being the cell index,
and the momentum representation
$\psi_p=(D_p, S_p, X_p, Y_p)$ 
of the TB wave function (when used as an index, the electron quasi-momentum vector is denoted by $p$).
Hence, the Schr\"odinger equation ${\rm i}\hbar{\rm d}_t {\hat \psi}_{p,\alpha} = [{\hat \psi}_{p,\alpha}, {\hat H}]$ for $\psi_{p,\alpha}(t) = \e^{-{\rm i}\eps t/\hbar}\psi_{p,\alpha},$ with $\alpha$ being the spin index $(\uparrow,\downarrow)$ (suppressed hereafter), takes the form
\eq{4ham}{
\fl \left( H_p^{(4\sigma)} -\eps\openone \right) \psi_p =
\left(
\ba{lccr}
   -\ed        & 0           & \tpd s_{_X}        & -\tpd s_{_Y} \\
   0            & -\ves        & \tsp s_{_X}        & \tsp s_{_Y} \\
   \tpd s_{_X}  & \tsp s_{_X} & -\ep               & -\tpp s_{_X}s_{_Y} \\
  -\tpd s_{_Y}  & \tsp s_{_Y} & -\tpp s_{_X}s_{_Y} & -\ep
\ea \right)
\left(\!\!
 \ba{c}
 D_p \\
 S_p\ \\
 X_p \\
 Y_p \\
 \ea \!\!
 \right) = 0,
}
where
\[
  \ed = \epsilon-\epsilon_d,\quad \ves = \epsilon-\epsilon_s,\quad \ep =
  \epsilon-\epsilon_p,
\]
and
\eqn{
s_{_X} & =   & 2\sin (\case{1}{2}p_x),\;  s_{_Y} = 2\sin (\case{1}{2}p_y),\;
x=\sin^2(\case{1}{2}p_x),\; y=\sin^2(\case{1}{2}p_y) \nn \\
0      & \le & p_x,\; p_y \le 2\pi . \nn
}
This $4\sigma$-band Hamiltonian is generic for the layered cuprates,
cf.~\Rref{OKA95}. We have also included the direct oxygen-oxygen exchange
$t_{pp}$ dominated by the $\sigma$ amplitude. The secular equation
\eq{seceq}{
\det\left( H_p^{(4\sigma)} - \epsilon\openone\right) =
{\cal A}xy + {\cal B}(x+y) + {\cal C} = 0
}
gives the spectrum and the canonical form of the CEC with energy-dependent coefficients
\eqn{
{\cal A}(\epsilon) &=& 16(4\tpd^2\tsp^2 + 2 \tsp^2\tpp\ed - 2 \tpd^2\tpp\ves -
\tpp^2\ed\ves) \nn \\
{\cal B}(\epsilon) &=& -4\ep(\tsp^2\ed+\tpd^2\ves) \label{ABC} \\
{\cal C}(\epsilon) &=& \ed\ves\ep^2. \nn
}
Hence, the explicit CEC equation reads as 
\eq{py}{
p_y = \pm\arcsin\sqrt{y},\mbox{ if } 0\le y=-\frac{{\cal B}x+{\cal C}}{{\cal
      A}x+{\cal B}}\le 1.
}
This equation reproduces the rounded square-shaped FS, centered at the
$(\pi,\pi)$ point, inherent for all layered cuprates. The best fit is achieved
when $\cal A,$ $\cal B$ and $\cal C$ are considered as fitting
parameters. Thus, for a CEC passing through the ${\sf D} = (p_{\rm d},p_{\rm
d})$ and ${\sf C} = (p_{\rm c},\pi)$ reference points, as indicated
in~\fref{fig:2}, the fitting coefficients (distinguished by the subscript $f$) in the canonical equation ${\cal A}_fxy +
{\cal B}_f(x+y) + {\cal C}_f = 0$ have the form
\eq{ABCfit}{
\ba{ll}
{\cal A}_f = 2x_{\rm d}-x_{\rm c} -1, & x_{\rm d}= \sin^2(p_{\rm d}/2) \\
{\cal B}_f = x_{\rm c}-x_{\rm d}^2,    & x_{\rm c}= \sin^2(p_{\rm c}/2) \\
{\cal C}_f = x_{\rm d}^2(x_{\rm c}+1)-2x_{\rm c}x_{\rm d},
\ea
}
and the resulting LCAO Fermi contour is quite compatible with the LDA
calculations for $\NCCO$~\cite{Massidda89,Yu91pr}. Due to the simple shape of
the FS the curves just coincide. We note also that the canonical
equation~(\ref{seceq}) would formally correspond to 1-band TB Hamiltonian of a
2D square lattice of the form
\[
  \epsilon({\bf p}) = -2t(\cos p_x +\cos p_y)+4t'\cos p_x\cos p_y,
\]
with strong energy dependence of the hopping parameters, 
where $t'$ is the anti-bonding hopping between the sites along the diagonal,
cf. references~\cite{Yu91,Norman}.

\subsection{Effective Hamiltonians}

Studies of the electronic structure of the layered cuprates have unambiguously
proved the existence of a large hole pocket---a rounded square centred at
the $(\pi,\pi)$ point. This observation is indicative for a Fermi level located
in a single band of dominant $\cud$ character. To address this band and the
related wave functions it is therefore convenient an effective Cu-Hamiltonian to
be derived by L\"owdin downfolding of the oxygen orbitals. This is equivalent to
expressing the oxygen amplitudes from the third and fourth rows of~\eref{4ham}
\eq{X,Y}{
\ba{l}
X=\frac{1}{\eta_p} \left[ \tpd s_{_X}\left(
  1+\frac{\tpp}{\ep}s_{_Y}^2 \right) D + \tsp s_{_X}\left(
  1-\frac{\tpp}{\ep}s_{_Y}^2 \right)S \right] \\
\\
Y=\frac{1}{\eta_p}\left[ -\tpd s_{_Y}\left(
   1+\frac{\tpp}{\ep}s_{_X}^2\right)D + \tsp s_{_Y}\left(
   1-\frac{\tpp}{\ep}s_{_X}^2\right)S \right],
\ea
}
where $\eta_p=\ep-\frac{\tpp^2}{\ep}s_{_X}^2s_{_Y}^2,$
and substituting back into the first and the second rows of the same equation.
Such a downfolding procedure results in the following energy-dependent copper 
Hamiltonian
\eq{Hcu}{
\fl H_{\rm Cu}(\epsilon)=
\left(
\ba{lr}
\epsilon_d + \frac{(2\tpd)^2}{\eta_p}\left( x+y
      +\frac{8\tpp}{\ep}xy \right) &
\frac{(2\tpd)(2\tsp)}{\eta_p}\left( x-y \right)\\
\\
\frac{(2\tpd)(2\tsp)}{\eta_p}\left( x-y \right) &
\epsilon_s + \frac{(2\tpd)^2}{\eta_p}\left( x+y
      -\frac{8\tpp}{\ep}xy \right)
\ea
\right),
}
which enters the effective Schr\"odinger equation $H_{\rm Cu}{D \choose
S}=\epsilon{D \choose S}.$ Thus, from \eref{X,Y} and \eref{Hcu} one can easily
obtain an approximate expression for the eigenvector corresponding
to a dominant $\cud$ character. Taking $D\approx 1,$ in the lowest order with
respect to the hopping amplitudes $t_{ll'}$ one has
\eq{3dvec}{
|\cud\rangle = \left(\! \ba{c}D\\S\\X\\Y \ea\!\right) \approx
                \left(\!\! \ba{c}
1                                       \\
(\tsp\tpd/\ves\ep)(s_{_X}^2 - s_{_Y}^2) \\
(\tpd/\eta_p)s_{_X}                     \\
-(\tpd/\eta_p)s_{_Y}
               \ea  \!\!\right),
}
i.e. $|X|^2+|Y|^2+|S|^2 \ll |D|^2\approx 1.$
We note that within this Cu scenario the Fermi level location and the CEC shape
are not sensitive to the $\tpp$ parameter. Therefore one can neglect the
oxygen-oxygen hopping as was done, for example,  by Andersen
\etal~\cite{OKA94,OKA95} (the importance of the $\tpp$ parameter has been considered by Markiewicz~\cite{Markiewicz}) and the band structure of the Hamiltonian~(\ref{Hcu})
for the same set of energy parameters as used in~\Rref{OKA95} is shown
in~\fref{fig:3} ({\it a}). In this case the FS can be fitted by its diagonal
alone, i.e. using only {\sf D} as a reference point. Hence an equation for
the Fermi energy follows, ${\cal A}(\eps_{_{\rm F}})x_{d}^{2}+2{\cal
B}(\eps_{_{\rm F}})x_d +{\cal C}(\eps_{_{\rm F}})=0,$ which yields
$\eps_{_{\rm F}}=2.5\mbox{ eV}$. As seen in~\fref{fig:3}~({\it b}), the deviation
from the
two-parametr fit, discussed in~\sref{sec:A} is almost vanishing thus
justifying the neglect of $\tpp$ and using one-parametr fit.

However, despite the excellent agreement between the LDA calculations, the LCAO
fit and the ARPES data regarding the FS shape, the theoretically calculated
conduction band width $w_{\rm c}$ in the layered cuprates is overestimated by
a factor of
2 or even 3~\cite{King}. Such a discrepancy may well point to some alternative
interpretations of the available experimental data.
In the following section we shall consider the possibility for a Fermi level
lying in an oxygen band.

\subsubsection{Oxygen scenario: the Abrikosov-Falkovsky model}
\label{AFM}

Various hints currently exist in favor of O$2p$ character of the states
near the Fermi level~\cite{oxgn-hints,Blackstead}. We consider that
these arguments cannot be {\it a priori} ignored. This is best see if, following
Abrikosov and Falkovsky~\cite{Abrikosov}, the experimental data are interpreted
within an alternative oxygen scenario.

Accordingly, the oxygen $2p$ level is assumed to lie above the $\cud$
level, and the Fermi level to fall into the upper oxygen band,
$\eps_d<\eps_p < \eps_{_{\rm F}} < \eps_s.$ The $\cud$ band is completely
filled in the metallic phase and the holes are found to be in the approximately
half-filled O$2p\sigma$ bands. To inspect such a possibility in detail we use
again the L\"owdin downfolding procedure now applied to Cu orbitals. From the
first and second rows of \eref{4ham} we express the copper amplitudes
\eq{D,S}{ \ba{l}
D = \frac{\tpd}{\ed}(s_{_X}X-s_{_Y}Y) \\ \\
S = \frac{\tsp}{\ves}(s_{_X}X+s_{_Y}Y)
\ea
}
and substitute them in the third and the fourth rows. This leads to an effective
oxygen Hamiltonian of the form
\eq{Hoxgn}{
H_{\rm O}(\epsilon)= B \left(\!\ba{lr}
                             s_{_X}s_{_X} &  s_{_X}s_{_Y}\\
                             s_{_Y}s_{_X} &  s_{_Y}s_{_Y}
                               \ea\!\right)
- t_{\rm eff}\left(\!\ba{lr}
                    0            &  s_{_X}s_{_Y}\\
                    s_{_Y}s_{_X} &  0
                     \ea\!\right)
}
with spectrum
\eq{ecb}{
\epsilon({\bf p}) = 2B(\eps)(x+y)\left[-1\pm \sqrt{ 1 +
                    \left(2\tau+\tau^2\right)
                    \frac{4xy}{(x+y)^2}} \, \right],
}
where
\eqn{
  B(\epsilon) = -\frac{\tpd^2}{\ed}+\frac{\tsp^2}{(-\ves)},\quad
  t_{\rm eff}(\epsilon) = \tpp + 2\frac{\tpd^2}{\ed}, \quad
  \tau(\epsilon) = t_{\rm eff}/B\\
  -\ves,\;\ed > 0
}
and the conduction band dispersion rate $\eps_{\rm c}({\bf
p})$ corresponds to the "+" sign for $|\tau |<1$. It should be noted
that~\eref{ecb} is an
exact result within the adopted $4\sigma$-band model. As a consequence, it is
easily realised that along the  $(0,0)$-$(\pi,0)$ direction the conduction band
is dispersionless, $\eps_{\rm c}(p_x,0)=0.$ This corresponds to the extended Van Hove
singularity observed in the ARPES experiment~\cite{Gofron} and we consider it
being a hint in favour of the oxygen scenario (the copper model would
give instead the usual Van Hove scenario).

Depending on the $\tau$ value two different limit cases occur. For $\tau\ll 1$
one gets a simple Pad\`e approximant
\eq{ec1}{
 \epsilon_{\rm c}({\bf p}) = 4t_{\rm eff}(\epsilon_{\rm c})\frac{2xy}{x+y}
}
and eigenvector of $H^{(4\sigma)}$
\eq{c1}{
 |{\rm c}\rangle = \left(\! \ba{c}D\\S\\X\\Y \ea\!\right) \approx
\frac{1}{\sqrt{s_{_X}^{2} + s_{_Y}^{2}}}\!\!
  \left(\!\ba{c}
         \frac{2\tpd}{\ed}s_{_X}s_{_Y}\\
         0     \\
        -s_{_Y}\\
         s_{_X}
  \ea\!\right),
}
normalized according to the inequality $|D|^2+|S|^2\ll |X|^2+|Y|^2\approx 1.$
This limit case acceptably describes the experimental ARPES data e.g. for
$\NCCO,$ material with single $\cuo$ planes and no other complicating
structural details. Schematic representation of the energy surface defined by~\eref{ec1}
is shown in~\fref{ARPES} ({\it a}).  
In~\fref{ARPES} ({\it b}) we have presented a comparison between the
ARPES data from~\Rref{King} and the Fermi contour calculated
according to~\eref{ec1} for $x=0.15.$ Note that {\it no fitting parameters} are
used and this contour should be referred to as an {\it ab initio} calculation
of the FS. 

The opposite limit case $t_{\rm eff} \gg B,$ i.e. $\tau\gg 1,$ has been analysed in
detail by Abrikosov and Falkovsky~\cite{Abrikosov}. The conduction
band dispersion rate $\epsilon_{\rm c}$ and the corresponding eigenvector of
the Hamiltonian $H_{\rm O}$~(\ref{Hoxgn}) now take the form
\eqn{
 \epsilon_{\rm c}({\bf p}) & = & 4t_{\rm eff}(\eps_{\rm c})\sqrt{xy}
\label{ec2}\\  |{\rm c}\rangle           & \approx & \frac{1}{\sqrt{2}}\!\!
                 \left(\!\ba{c}
                         (\tpd/\ed)(s_{_X}+s_{_Y})\\
                         (\tsp/\ves)(s_{_X}-s_{_Y})\\
                         1\\
                        -1
                         \ea\!\right),
\label{c2}
}
provided that $|D|^2+|S|^2\ll |X|^2+|Y|^2\approx 1.$
In other words, the last approximation, $\tau\gg 1,$ corresponds to a pure
oxygen model where only hoppings between oxygen ions are taken into account.
Clearly, this model is the complementary to the copper scenario and is based on
an effect completely neglected in its copper "counterpart", where $\tpp \equiv 0.$ 
This limit case of the oxygen scenario suitably describes the ARUPS
experimental data for
Pb$_{0.42}$Bi$_{1.73}$Sr$_{1.94}$Ca$_{1.3}$Cu$_{1.92}$O$_{8+x}$~\cite{Aebi}.
The FS of the latter is fitted by its diagonal (the {\sf D} point) according
to the Abrikosov-Falkovsky relation~(\ref{ec2}) and the result is shown
in~\fref{fig:4}. 

There exist a tremendous number of ARPES/ARUPS data for layered cuprates which makes the reviewing of all those spectra impossible. To illustrate our TB model we have chosen data for the Pb substitution for Bi in Bi$_2$Sr$_2$CaCu$_2$O$_8,$ see \fref{fig:4}. In this case the $\cuo$ planes are quite flat and the ARPES data are not distorted by structural details. When present, distortions were misinterpreted as a manifestation of strong antiferromagnetic correlations.  We believe, however, that the experiment by Aebi \etal~\cite{Aebi} reveals the main feature of the $\cuo$ plane band structure---the large hole pocket found to be in agreement with the one-particle band calculations.

Besides the good agreement between the theory and the experiment, regarding the
FS shape, we should also point out the compatibility between the calculated and
the experimental conduction bandwidth. Indeed, within the Abrikosov-Falkovsky
model~\cite{Abrikosov}, according to~\eref{ec2}, one gets for the conduction
bandwidth $0\le \varepsilon_{\rm c}({\bf p}) \le w_{\rm c} \approx 4\tpp,$
which coincides with the value obtained from~\eref{ec1} provided that $\tpd^2\ll
\tpp(\eps_{_{\rm F}}-\eps_d).$
The {\it ab initio} calculation of $\tpp$ as a surface integral (see~\ref{ap1}), 
making use of atomic wave functions standard for the
quantum mechanical calculations, gives $\tpp \approx$~200--350~meV
in different estimations. This range is in acceptable agreement with the
experimental $w_{\rm c}\simeq 1\mbox{ eV}$~\cite{King}; within the LCAO model
an exact analytic result for $w_{\rm c}$ can be obtained from the equation
$w_{\rm c} = 4\tpp + 8\tpd^2/(w_{\rm c}-\epsilon_d).$

We note also that the TB analysis allows the bands to be unambiguously
classified with respect to the atomic levels from which they arise. Within such
terms, for the oxygen scenario one can describe the metal$\rightarrow$insulator
transition as being the charge transfer $\rm
Cu^{1+}O^{1\frac{1}{2}-}_{2}\rightarrow Cu^{2+}O_{2}^{2-}.$ The possibility for
monovalent copper Cu$^{1+}$ in the superconducting state is discussed, for
example, by Romberg \etal~\cite{Romberg}.


\section{Conduction bands of RuO$_2$ plane}
\label{sec:III}

Sr$_2$RuO$_4$ is the first coper-free perovskite superconductor isostructural to the
high-$T_c$ cuprates~\cite{Maeno}. The layered ruthenates, just like the layered
cuprates, are strongly anisotropic and in a first approximation the nature of the
conduction band(s) can be understood by analysing the bare $\ruo$ plane. One
should repeat the same steps as in the previous section but now having Ru
instead of Cu and the Fermi level located in the metallic bands of Ru$4d\pi$
character. To be specific, the conduction bands arise form the hybridisation
between the Ru$4d_{xy},$ Ru$4d_{yz},$ Ru$4d_{zx}$ and O$_{\rm a}2p_y,$ O$_{\rm
b}2p_x$, O$_{\rm a,b}2p_z$ $\pi$-orbitals. The LCAO wave function spanned over
the four perpendicular to the $\ruo$ plane orbitals reads as
\eqn{
\fl \Psi_{_{\rm LCAO}}^{(z)}( {\bf r} ) = \frac{1}{\sqrt{N}}\sum_{\bf
p}\sum_{\bf n} \left[
D_{zx,{\bf n}}\psi_{_{ {\rm Ru}4d_{zx}}} ({\bf r}-a_{_0}{\bf n}) +
D_{zy,{\bf n}}\psi_{_{ {\rm Ru}4d_{zy}}} ({\bf r}-a_{_0}{\bf n}) \label{psiz}
\right. \\
+ \left. \e^{{\rm i}\varphi_a}Z_{a,{\bf n}}\psi_{_{ {\rm O_a} 2p_z}} ({\bf r}-{\bf
R}_{\rm O_a}-a_{_0}{\bf n})
+ \e^{{\rm i}\varphi_b}Z_{b,{\bf n}}\psi_{_{ {\rm O_b} 2p_z}} ({\bf r}-{\bf  R}_{\rm
O_b}-a_{_0}{\bf n}) \right] \e^{{\rm i}\bf p\cdot n},\nn
}
Hence, the $\pi$-analog of~\eref{4ham} takes the form
\eq{ZSchr}{
\fl \left( H_{p}^{(z)} - \eps\openone\right)\psi_{p}^{(z)} =
        \left(\! \ba{lccr}
-\varepsilon_{zx}&                 0 & \tzzx s_{_X}     & 0 \\
0                & -\varepsilon_{zy} & 0                & t_{z,zy} s_{_Y} \\
\tzzx s_{_X}     &          0        & -\varepsilon_{za}& -\tzz c_{_X}c_{_Y} \\
0                & t_{z,zy} s_{_Y}   & -\tzz c_{_X}c_{_Y}  & -\varepsilon_{zb}
                   \ea\! \right)
\left(\!\!\ba{c} D_{zx}\\D_{zy}\\Z_a\\Z_b \ea\!\!\right) = 0,
}
where
\eq{eps}{ \ba{lll}
\ezx             = \eps - \eps_{zx},& \varepsilon_{za} = \eps - \eps_{za}, & c_{_X} = 2\cos(p_x/2), \\
\varepsilon_{zy} = \eps - \eps_{zy},& \varepsilon_{zb} = \eps - \eps_{zb}, & c_{_Y} = 2\cos(p_y/2), 
\ea
}
and $\epsilon_{zx},$ $\epsilon_{zy},$ $\epsilon_{za},$ and $\epsilon_{zb}$ are
the single site energies respectively for Ru$4d_{zx},$ Ru$4d_{zy}$ and
O$_{\rm a}2p_z,$ O$_{\rm b}2p_z$ orbitals. $\tzz$ stands for the hopping
between the latter two orbitals and, if a negligible orthorhombic distortion is
assumed, the metal-oxygen $\pi$-hopping parameters are equal, $t_{z,zy}=\tzzx$
and also $\eps_z = \eps_{za}=\eps_{zb}$. The phase factors
$\e^{{\rm i}\varphi_{a,b}}$ in~\eref{psiz} are chosen in compliance
with~\Rref{OKA95}, see equation~\eref{Blochwf}. 

Identically, writing the LCAO wave function spanned over the three in-plane
$\pi$-orbitals Ru$4d_{xy},$ O$_{\rm a}2p_y,$ and O$_{\rm b}2p_x$ in the
way in which~\eref{psiz} is designed one has for the "in-plane" Schr\"odinger equation
\eq{XYSchr}{
\fl \left( H_{p}^{(xy)} - \eps\openone\right)\psi_{p}^{(xy)} =
        \left(\! \ba{lcr}
-\exy       & \tpdp s_{_X}       & \tpdp s_{_Y}       \\
\tpdp s_{_X}     & -\varepsilon_{ya}         & \tpp' s_{_X}s_{_Y} \\
\tpdp s_{_Y}     & \tpp' s_{_X}s_{_Y} & -\varepsilon_{xb}
                   \ea\! \right)
\left(\!\!\ba{c} D_{xy}\\Y_a\\X_b \ea\!\!\right) = 0,
}
where $\tpdp$ denotes the hopping Ru$4d_{xy}\rightarrow {\rm O_{a,b}}2p\pi$ and
$\tpp',$ respectively,${\rm O_a}2p_y \rightarrow {\rm O_b}2p_x.$ The
definitions for the other energy parameters are in analogy to~\eref{eps} (for
negligible orthorhombic distortion $\eps_{ya}=\eps_{xb}\neq \eps_z$). Thus, the
$\pi$-Hamiltonian of the $\ruo$ plane takes the form
\eq{7ham}{
  H^{(\pi)}= \sum_{p,\alpha=\uparrow\!,\downarrow}
\psi_{p,\alpha}^{(z)\dagger} H_p^{(z)} \psi_{p,\alpha}^{(z)} +
\psi_{p,\alpha}^{(xy)\dagger} H_p^{(xy)} \psi_{p,\alpha}^{(xy)}.
}
In a previous paper~\cite{JLTP} we have derived the corresponding secular
equations and now we shall only provide the final expressions in terms of the
notations used here
\[
  \det(H^{(z,xy)}_{p}-\eps\openone) = {\cal A}^{(z,xy)}xy+{\cal
B}^{(z,xy)}(x+y)+{\cal C}^{(z,xy)}= 0,
\]
\eq{coeff}{ \ba{ll}
{\cal A}^{(z)} = 16(\tzzx^4-\tzz^2\ezx^2) &
{\cal A}^{(xy)} = 32\tpp'\tpdp^2-16\exy\tpp'^2 \\
{\cal B}^{(z)} = -16\tzz^2\ezx^2-4\tzzx^2\ezx\varepsilon_z, &
{\cal B}^{(xy)} = -\tpdp^2\varepsilon_{ya} \\
{\cal C}^{(z)} = \ezx^2(\varepsilon_{z}^{2}-16\tzz^2) &
{\cal C}^{(xy)} = \exy\varepsilon_{ya}^{2}
\ea .
}
The three sheets of the Fermi surface in Sr$_2$RuO$_4$ fitted to the ARPES
data given by Lu \etal~\cite{Lu96} are shown in~\fref{fig:5}~({\it b}). To determine the
Hamiltonian parameters we have made use of the dispersion rate values at the
high-symmetry points of the Brillouin zone. To the best of our knowledge, the TB analysis of the Sr$_2$RuO$_4$ band structure was first performed in \Rref{JLTP} (subsequently, the latter results were reproduced in \Rref{Noce} without referring to \Rref{JLTP}). The $\ruo$-plane band structure resulting from the set of parameters
\eq{params}{\ba{lll}
 \tzz=\tpp'= 0.3\mbox{ eV},& \varepsilon_{z}= -2.3\mbox{ eV}, & \exy=
 -1.62\mbox{   eV},\\ \\
 \tpdp=\tzzx = 1\mbox{ eV},& \ezx = -1.3\mbox{ eV}, &
 \varepsilon_{ya,xb}=-2.62\mbox{  eV}. \ea
}
is shown in~\fref{fig:5}~({\it a}). This fit is subjected to the requirement of providing as
good as possible a description of the narrow energy interval around $\eps_{_{\rm F}}$
whereas the filled bands far below the Fermi level match only qualitatively to the LDA
calculations by Oguchi~\cite{Oguchi} and Singh~\cite{Singh}. In addition we
note that the de Haas-van Alphen (dHvA) measurements~\cite{Makenzie} of the
Sr$_2$RuO$_4$ FS differ from the ARPES results~\cite{Lu96}. Thus, fitting the
dHvA data by using modified TB parameters is a natural refinement of the
proposed model. We note that the diamond-shaped hole pocket, centred at the X point (see \fref{fig:5}~({\it b})), is very sensitive to the `game of parameters'. For that band the van Hove energy is fairly close to the Fermi energy. As a result, a minor change in the parameters could drive a van Hove transition transforming this hole pocket to an electron one, centred at the $\Gamma$ point. Indeed, such a band configuration has been  recently observed also in the ARPES revision of the Sr$_2$RuO$_4$ Fermi surface~\cite{Puchkov}. This can be easily traced already from the energy surfaces $\epsilon({\bf p})$ calculated earlier in~\Rref{JLTP}. The comparison of the ARPES data with TB energy surfaces could be a subject of a separate study.


\section{Discussion}
\label{sec:IV}

The LCAO analysis of the layered perovskites band structure, performed in the
preceding sections, manifests a good compatibility with the experimental
data and the band calculations as well. Due to the strong anisotropy of these
materials, their FS within a reasonable approximation is determined by the
properties of the bare $\cuo$ or $\ruo$ planes.

Despite these planes having identical crystal structure, their electronic structures
are quite different. While for the $\ruo$ plane the Fermi level crosses metallic
$\pi$-bands, the conduction band of the $\cuo$ plane is described by a
$\sigma$-Hamiltonian~\eref{4ham}. The latter gives for the $\cuo$ plane a
large hole pocket centered at the $(\pi,\pi)$ point. Its shape, if no additional
sheets exist, is well described by the exact analytic results within the LCAO
model, \Eref{seceq}, as found for $\NCCO$~\cite{Yu91pr,King} and
Pb$_{0.42}$Bi$_{1.73}$Sr$_{1.94}$Ca$_{1.3}$Cu$_{1.92}$O$_{8+x}$~\cite{Aebi}.
For a number of other cuprates,
YBa$_2$Cu$_3$O$_{7-\delta}$~\cite{Yu87},
YBa$_2$Cu$_4$O$_8$~\cite{Gofron},
Bi$_2$Sr$_2$CaCu$_2$O$_8$~\cite{Marshall95,MYF88},
Bi$_2$Sr$_2$CuO$_6$~\cite{SP95},
the infinite-layered superconductor Sr$_{1-x}$Ca$_x$CuO$_2$~\cite{NG93},
HgBa$_2$Ca$_2$Cu$_3$O$_{8+\delta}$~\cite{Singh94},
HgBa$_2$CuO$_{4+\delta}$~\cite{Novikov},
HgBa$_2$Ca$_{n-1}$Cu$_n$O$_{2n+2+\delta}$~\cite{NF93},
Tl$_2$Ba$_2$Ca$_{n-1}$Cu$_n$O$_{4+2n},$~\cite{Pickett},
Sr$_2$CuO$_2$F$_2$,
Sr$_2$CuO$_2$Cl$_2$,
Ca$_2$CuO$_2$Cl$_2$~\cite{Novikov95},
this large hole pocket is easily identified. For all of the above compounds,
however, its shape is usually deformed due to appearance of additional sheets of
the Fermi surface originating from accessoires of the crystal structure.

As the most important implication for the $\cuo$ plane we should point out the
intrinsic alternative about the Fermi level location (see \sref{sec:II}). It is
commonly believed that the states at the FS are of dominant $\cud$ character
(see e.g.~\Rref{OKA95}). Nevertheless, the spectroscopic data for the FS can be
equally well interpreted within the oxygen scenario, according to which the FS
states are of dominant O$2p\sigma$ character. A number of indications exist in favour
of the oxygen model and the importance of the $\tpp$ hopping
amplitude~\cite{Markiewicz,oxgn-hints}: 
\begin{itemize}
\item[({\it{}i})] ${\rm O}1s \rightarrow {\rm O}2p$ transitions  observed in EELS experiments for the metallic phase of the layered cuprates, which reveal an unfilled O$2p$ atomic shell; 
\item[({\it{}ii})] the oxygen scenario reproduces in a natural way the extended van Hove singularity observed in the ARPES experiments while the Cu scenario fails to describe it; 
\item[({\it{}iii})] the metal-insulator transition can be easily described; 
\item[({\it{}iv})] the width of the conduction band is directly related to the atomic wave functions.
\end{itemize}
Some authors even "wager that the oxygen model will
win"~\cite{Blackstead} (if the oxygen scenario is
corroborated, due to the cancellation of the largest amplitude $\tsp$ the small
hoppings $\tpd$ and $\tpp$ should be properly evaluated eventually as surface
integrals (see~\ref{ap1}) and some band calculations may well need a
revision).  
It would be quite valuable if a muffin-tin calculation of H$_{2}^{+}$
ion was performed and compared with the exact results when the
hopping integral is comparatively small, of the order of the one that fits the
ARPES data $\tpp\sim 200\mbox{ meV}.$ 
We also note that even the copper model
gives an estimation for $\tpp$ closer to the experiment than the LDA
calculations. 
The smallness of $\tpp$ within the oxygen scenario, on the other hand, is guaranteed by the nonbonding character of the conduction band. This scenario, therefore, can easily display heavy fermion behaviour, i.e. effective mass $m_{\rm eff}\stackrel{\tpp\rightarrow 0}{\longrightarrow}\mbox{huge},$ and density of states (DOS) $\propto m_{\rm eff}\propto 1/\tpp$ (we note that no realistic band calculations for heavy fermion systems can be performed without employing the asymptotic methods from the atomic physics). It is also instructive to compare the TB analyses of heavy fermion systems and layered cuprates. 
The alternative for the Fermi level location (metallic {\it vs.} oxygen band) exist for the cubic bismuthates as well~\cite{cubic,ACAR}. When the Fermi level falls into heavy fermion oxygen bands, one of the isoenergy surfaces is a rounded cube~\cite{cubic}. Indeed, such a isoenergy surface has been recently confirmed by the LMTO method applied to Ba$_{0.6}$K$_{0.4}$BiO$_3$~\cite{Savrazov}.  

Due to the equally good fit of the results for the FS of the layered cuprates  within
the two models we can infer that at present any final judgement about
this alternative would be premature. Thus far we consider that the oxygen model
should be taken into account in the interpretation of the experimental data.
Moreover, the angular dependence of the superconducting order parameter
$\Delta({\bf p}) \propto \cos(p_x) - \cos(p_y)$ is readily derived within the
standard BCS treatment of the oxygen-oxygen superexchange~\cite{Mishonov}.
Analysis of some extra spectroscopic data by means of different models would
finally solve this dilemma. This cannot be done within the framework of the TB method.
A coherent picture requires a thorough study, where the TB model is just an useful tool to test the properties of a given solution. 

Up to now, the applicability of the LCAO approximation to the electron
structure of the layered cuprates can be considered as being proved. The basis
function of the LCAO Hamiltonian can be included in a realistic one-electron
part of the lattice Hamiltonians for the layered perovskites. This is an
indispensable step preceding the inclusion of the electron-electron
superexchange, electron-phonon interaction or any other kind of interaction
between conducting electrons.

\ack

The authors are especially thankful to  P~Aebi for being so kind to
provide them with extra details on ARUPS spectra as well as for the
correspondence on this topic. 
We are much indebted to J~Indekeu for the hospitality and 
good atmosphere during completion of this work, and would like to thank 
R Danev, I Genchev and R Koleva for the collaboration in the initial stages of this study. 
This paper was partially supported by the Bulgarian NSF No.~627/1996, 
the Belgian DWTC, the Flemish Government Programme VIS/97/01, the IUAP and the GOA.

\appendix
\section{Calculation of O-O hopping amplitude by the surface integral method}
\label{ap1}

From quantum mechanics~\cite{LL3} it is well known that the usual for the
quantum chemistry calculation of the hopping integrals as matrix elements of
the single particle Hamiltonian does not work when the overlap between the
atomic functions is too weak. If the hopping integrals are much smaller than
the detachment energy, they should be calculated as surface integrals using
(eventually distorted by the polarisation) atomic wave functions.

Such an approach has been applied by Landau and Lifshitz~\cite{LL3} and
Herring and Flicker~\cite{Herring} to the simple H$^{+}_{2}$ problem and now the asymptotic
methods are well developed in the physics of atomic collisions~\cite{Smirnov}.
On the basis of the above problem one can easily verify that the atomic sphere muffin-tin approximation of the Coulomb
potentials usual for the
condensed matter physics undergoes {\it fiasco} when the hopping integrals are of the order of
200--300 meV. Therefore, the factor 2--3 misfit for a single electron
problem cannot be ascribed to the strong-correlation effects, renormalizations
and other incantations which are often used to account for the discrepancy
between the experimental bandwidth and the LDA calculations.

Usually condensed matter physics does not need asymptotically accurate methods
for calculation of hopping integrals which leads to zero overlap between
the muffin-tin and asymptotic methods. However, for the perovskites the largest
hopping $\tsp$ cancels in the expression for the upper oxygen band
$\eps_{\rm c}({\bf p}).$ Thus, small hoppings become essential, but having no
influence on the other bands, and the necessity of taking into account the $\tpp$
is of topological nature.

Following the calculations for H$^{+}_{2}$~\cite{LL3}, in a simplified picture
of two oxygen atoms O$_{\rm a},$ O$_{\rm b}$ separated by distance 
$d=\case{\sqrt{2}}{2}a_{_0}$ the surface integral method gives for 
the oxygen-oxygen exchange the following explicit expression
\eq{A1}{
\tpp = \frac{\hbar^2}{2m}\int\!\!\!\int\limits_{\cal S} (\psi\!_{_{\rm O_a}}\partial_z
      \psi\!_{_{\rm O_b}} - \psi\!_{_{\rm O_b}}\partial_z \psi\!_{_{\rm O_a}})
      {\rm d} x\,{\rm d} y,
}
where the integral is taken over the surface $\cal S$ halving $d,$ and $m$ is
the electron mass. Thus $\tpp=\tpp(\xi)$ is a function of $\xi = \kappa |{\bf
R}_{\rm O_a}-{\bf R}_{\rm O_b}|$ with $\kappa^2/2$ being the oxygen detachment
energy in atomic units and the detailed derivation of~\eref{A1} can be found,
for example, in~\Rref{Smirnov}.

We note that the derivation of $\tpp(\xi)$ imposes no restrictions on the basis
set $\{\psi\}$ used. Hence we choose $\{\psi\!_{_{\rm O_{a,b}}}\}$ to be the
simplest minimal (MINI) basis used~\cite{Huzinaga}, for example, in the GAMESS
package for doing {\it ab initio} electronic structure calculations~\cite{GAMESS}. The MINI
bases are three Gaussian expansions of each atomic orbital. The exponents and
contraction coefficients are optimised for each element, and the $s$ and $p$
exponents are not constrained to be equal.

Accordingly, the oxygen $2p$ radial wave function $R_{2p}(r)$ is replaced by a
Gaussian expansion $R^{(G)}_{2p}(r)$ and has the form
\eq{R2p}{
   R^{(G)}_{2p}(\zeta,{\bf r}) =
  \sum\limits_{i=1}^{3}C_{2p,i}g_{2p,i}(\zeta_{2p,i},{\bf r}),
}
where $g_{2p}(\zeta,{\bf r})= A_{2p,i}\e^{-\zeta_{2p,i} r^2},$ and the
coefficients for oxygen are given in~\tref{tab}. It is then normalized
to unity according to $\int_{0}^{\infty}R^{(G)2}_{2p}r^2{\rm d} r =1.$

By multiplying with the corresponding cubic harmonic the oxygen wave functions
are brought into the form
\eq{oxwf}{
 \psi_{_{\rm O_a}}({\bf r}_{\rm a})= R^{(G)}_{2p}(\zeta,{\bf r}_{\rm
a})\sqrt{\frac{3}{4\pi}}\frac{x_{\rm a}}{r_{\rm a}}, \quad
 \left\{ \ba{l}{\bf r}_{\rm a} = {\bf
r} - {\bf R}_{\rm O_a}\\ r_{\rm a} = |{\bf r}_{\rm a}| \ea \right. ,
}
and analogically for $\psi_{_{\rm O_b}}({\bf r} - {\bf R}_{\rm O_b}).$
Substituting~(\ref{oxwf}) in~(\ref{A1}) we get
\[
  \tpp^{^{\rm (MINI)}} = 340\; \mbox{meV}.
\]
In~\Rref{LT21} the same integral has been calculated with $\{\psi\}$ being
the asymptotic wave functions~\cite{Smirnov} appropriately tailored to the MINI
basis at their outermost inflection points $r^{(i)}$, i.e.
\eq{asymp}{
 R_{2p}(r) = \left\{ \ba{ll}
              R^{(G)}_{2p}(r),                        &  r \le r^{(i)}\\
              \\
              A\frac{\sqrt{2\kappa}}{r}e^{-\kappa r}, &  r\ge r^{(i)}
              \ea \right.
}
with $\kappa = 0.329$ and $A= 0.5.$ The value obtained is
\[
  \tpp^{\rm (asymp)} = 210\;\mbox{meV},
\]
found to be in good agreement with that fitted from the ARPES experiment within
the oxygen scenario. Similar calculation, for example, gives for the $\tpd$ and
$\tsp$ hoppings
\[
\tpd^{^{\rm (MINI)}} = 580\mbox{ meV,}\quad \tsp^{^{\rm (MINI)}} \sim 2.5\mbox{
eV}.
\]

\noindent
{\small
\textit{Note added in proof}.
In a very recent paper by Campuzano J C \etal 1999 [\textit{Phys. Rev.
Lett.} \textbf{83} 3709] the ARPES Fermi surface of pure
Bi$_2$Sr$_2$CaCu$_2$O$_{8+\delta}$ has been presented in the inset of
their figure~1 (a).
This experimental finding is in excellent agreement with our
tight-binding fit to the Fermi surface of 
Pb$_{0.42}$Bi$_{1.73}$Sr$_{1.94}$Ca$_{1.3}$Cu$_{1.92}$O$_{8+x}$,
studied by Schwaller P and co-workers in reference 6, 
given in figure~5 of the present paper.
The remarkable coincidence of the Fermi surfaces of these two
compounds is a nice confirmation that Pb substitution for Bi is irrelevant
for the band structure of the CuO$_2$ plane and the Fermi surface of the
latter is therefore revealed to be a common feature.
}


%
\Figures

\begin{figure}
\caption{
Schematic of a $\cuo$ plane (only orbitals relevant to the discussion are
depicted). The solid square represents the unit cell with respect to which the
positions of the other cells are determined. The indices of
the wave function amplitudes involved in the LCAO
Hamiltonian~(\protect{\ref{LCAOham}}) are given in brackets. 
The rules for determining the signs of hopping integrals $\tpd,$ $\tsp$, 
and $\tpp$ are shown as well.
}
\label{fig:1}
\end{figure}

\begin{figure}
\caption{
LDA Fermi contour of $\NCCO$ (dotted line) calculated by Yu and
Freeman~\protect{\cite{Yu91pr}} (with the kind permission of the authors), and the
LCAO fit (solid line) according to~\protect{\eref{seceq}}.
The fitting procedure uses ${\sf C}$ and ${\sf D}$ as reference points.
}
\label{fig:2}
\end{figure}

\begin{figure}
\caption{
({\it{}a}) Electron band structure of the generic for the $\cuo$ plane $4\sigma$-band
Hamiltonian using the parameters from~\protect{\Rref{OKA95}} and the Fermi level
$\eps_{_{\rm F}}=2.5\mbox{ eV}$ fitted from the LDA calculation by Yu and
Freeman~\protect{\cite{Yu91pr}};
({\it{}b}) The LCAO Fermi contour (solid line) fitted to the LDA Fermi surface
(dashed line) for $\NCCO$~\protect{\cite{Yu91pr}} using only {\sf D}
as a reference point. The deviation of the fit at the {\sf C} point is
negligible.}
\label{fig:3}
\end{figure}

\begin{figure}
\caption{
({\it{}a}) Energy dispersion of the nonbonding oxygen band $\epsilon_{\rm c}({\bf{}p})$, equation~\protect{\eref{ec1}}. A few cuts through the energy surface, i.e. CEC, are presented together with the dispersion along the high symmetry lines in the Brilloiun zone;
({\it{}b}) The Fermi surface of $\NCCO$ (solid line)
determined by equation~\protect{\eref{ec1}} for $x=0.15$ (shaded slice in panel ({\it{}a})) and compared with
experimental data (points with error bars) for the same value of $x$ after King
\etal~\protect{\cite{King}}. $\theta$ and $\varphi$ denote the polar and
azimuthal emission angles, respectively, measured in degrees. The empty dashed
circles show {\bf k}-space locations where ARPES experiments have been
performed (cf. figure~2 in~\protect{\Rref{King}}) and their diameter
corresponds to $2^{\circ}$ experimental resolution.
}
\label{ARPES}
\end{figure}

\begin{figure}
\caption{
({\it{}a}) ARUPS Fermi surface of
Pb$_{0.42}$Bi$_{1.73}$Sr$_{1.94}$Ca$_{1.3}$Cu$_{1.92}$O$_{8+x}$ by Aebi
\etal~\protect{\cite{Aebi}};
({\it{}b}) LCAO fit to ({\it{}a}) according to the Abrikosov-Falkovsky
model~\protect{\cite{Abrikosov}} using the {\sf D} reference point with
$p_{\rm d}=0.171\times 2\pi$.}
\label{fig:4}
\end{figure}

\begin{figure}
\caption{
({\it{}a}) LCAO band structure of Sr$_2$RuO$_4$ according to~\protect{\eref{7ham}}.
The Fermi level (dashed line) crosses the three Ru$4d\varepsilon$ bands of
the $\ruo$ plane;
({\it{}b}) LCAO fit (solid lines) to the ARPES data (circles) by Lu
\etal~\protect{\cite{Lu96}}, cf. also~\protect{\Rref{JLTP}}.}
\label{fig:5}
\end{figure}

\Tables

\begin{table}
\caption{Coefficients for the oxygen $2p$ wave function in the MINI
basis~\protect{\cite{GAMESS}}.}
\label{tab}
\end{table}

\begin{tabular}{llll} \br
$i$  & $C_{2p,i}$   & $A_{2p,i}$ & $\zeta_{2p,i}$ \\ \hline
1   & 8.2741400  & 2.485782 & 0.708520  \\
2   & 1.1715463  & 1.333720 & 0.476594  \\
3   & 0.3030130  & 0.263299 & 0.130440  \\
\br
\end{tabular}
\newpage
\begin{center}
\vspace*{2cm}
Figure 1.\\
\ \\
\epsfig{file=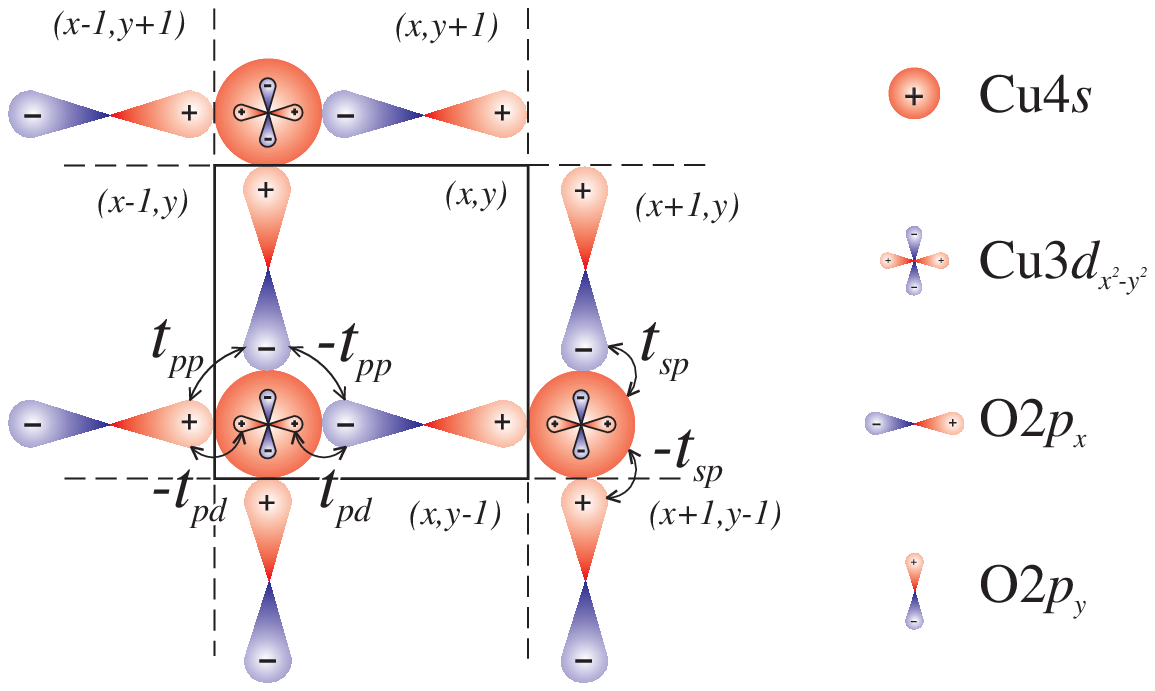,width=12cm}\\
\vskip3cm
Figure 2.\\
\ \\
\epsfig{file=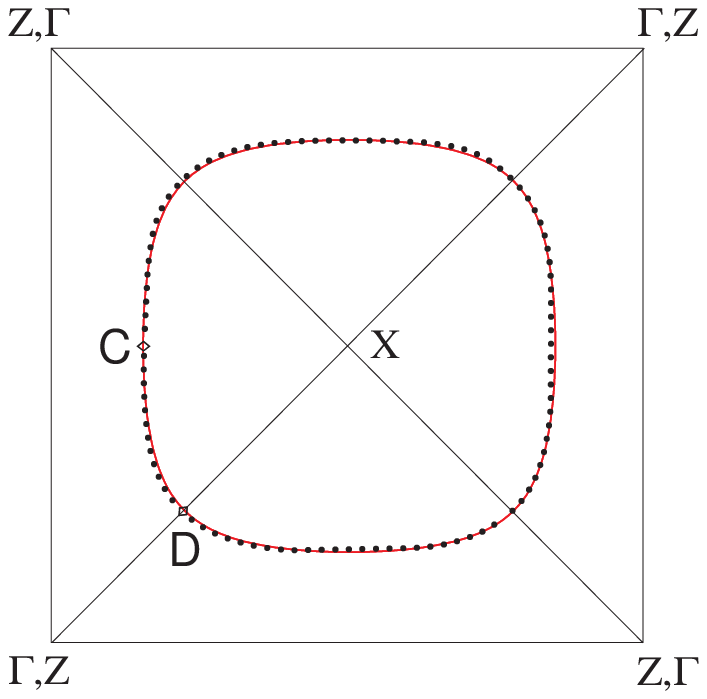,width=6cm}
\newpage
\vspace*{2cm}
Figure 3.\\
\ \\
\epsfig{file=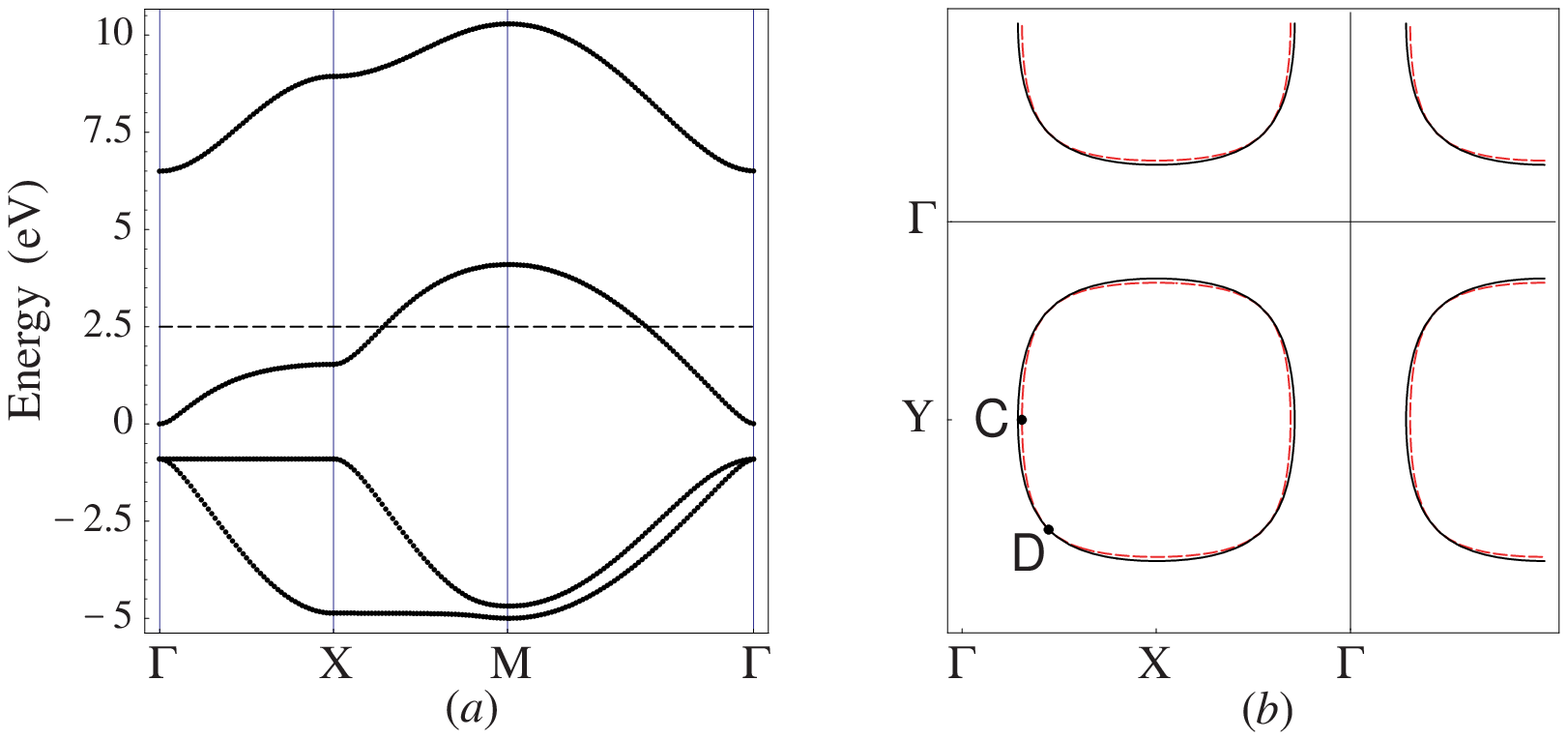,width=12cm}\\
\vskip3cm
Figure 4.\\
\ \\
\epsfig{file=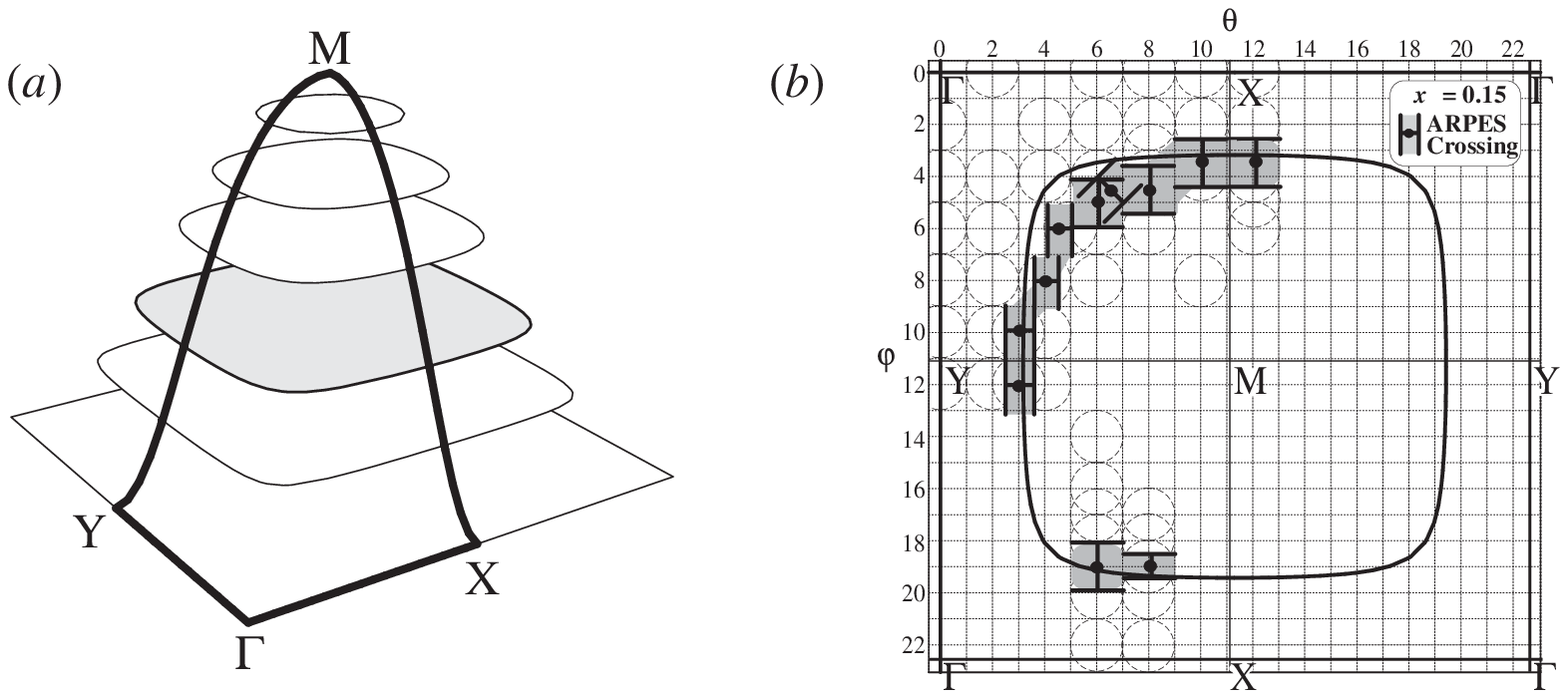,width=12cm}
\newpage
\vspace*{2cm}
Figure 5.\\
\ \\
\epsfig{file=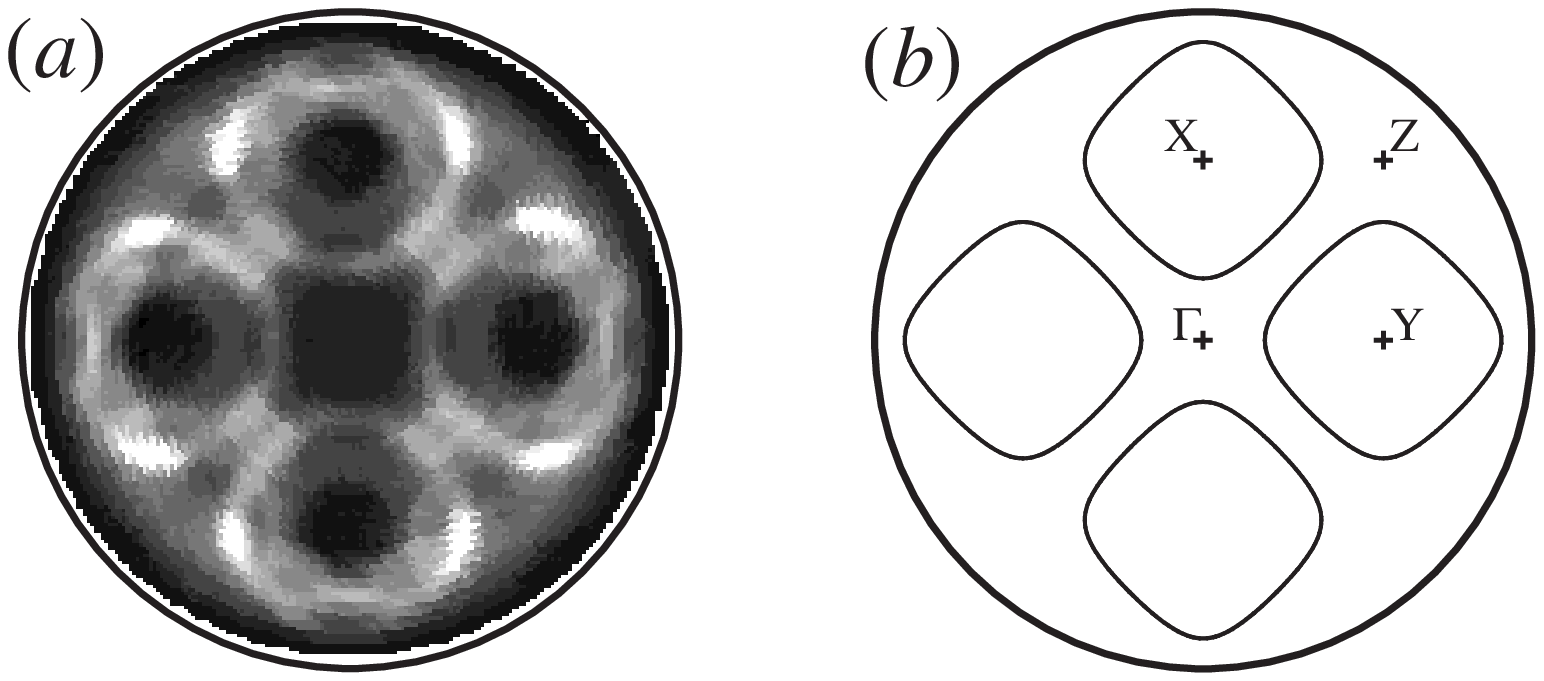,width=12cm}\\
\vskip3cm
Figure 6.\\
\ \\
\epsfig{file=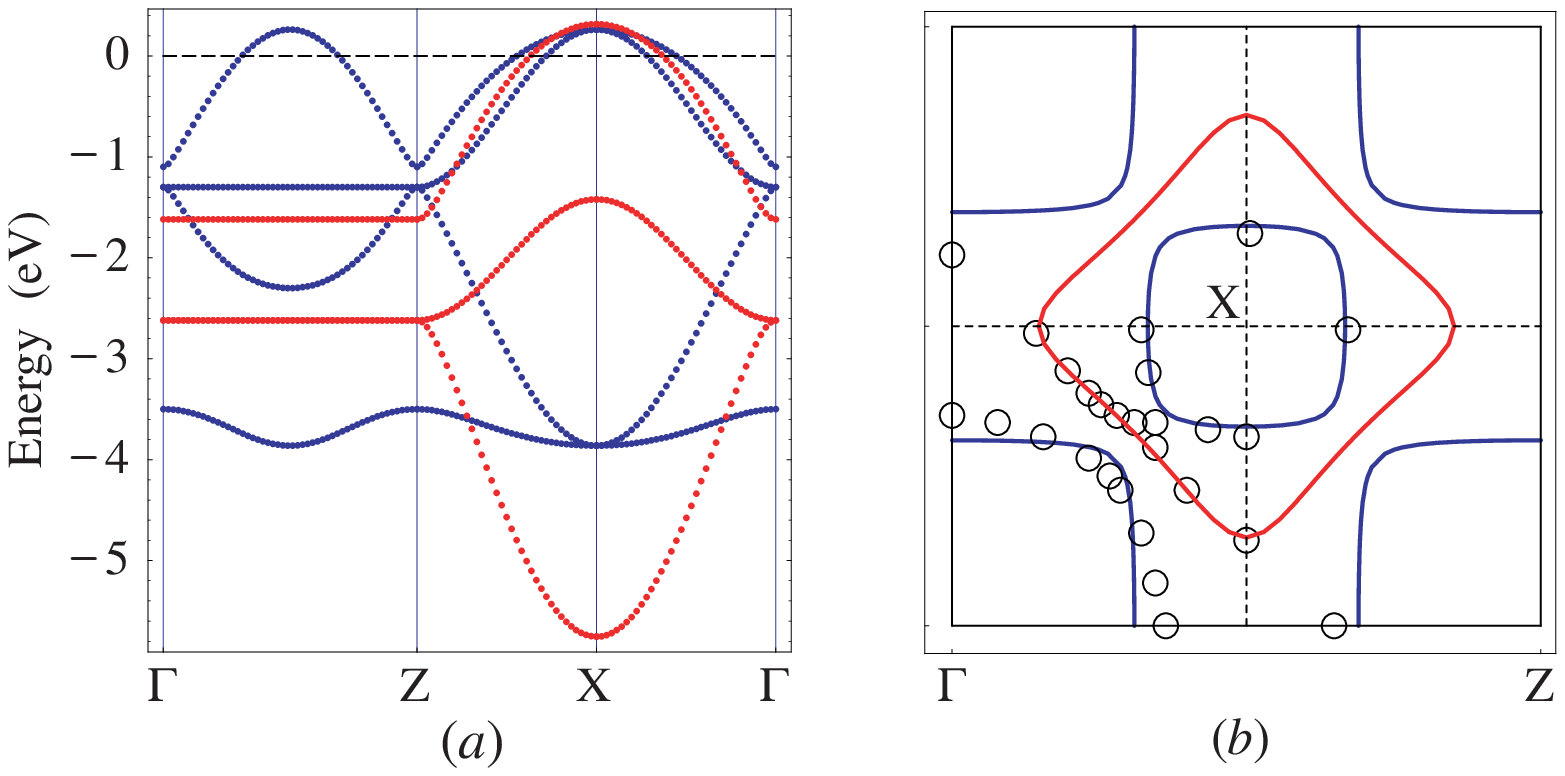,width=12cm}
\end{center}
\end{document}